\documentclass[10pt]{article}

\usepackage{amsmath}
\usepackage{amssymb}

\usepackage{graphicx}

\usepackage{cite}

\usepackage{color}

\makeatletter
\renewcommand{\@biblabel}[1]{\quad#1.}
\makeatother

\date{}

\begin{document}

\begin{flushleft}
{\Large
\textbf{Natural gaits of the non-pathological flat foot and high-arched foot}
}
\\
Yifang Fan$^{1,\ast}$,
Yubo Fan$^{2,\ast}$,
Zhiyu Li$^{3}$,
Changsheng Lv$^{1}$,
Donglin Luo$^{1}$
\\
\bf{1} Center for Scientific Research, Guangzhou Institute of Physical Education, Guangzhou 510500, P.R. China
\\
\bf{2} Key Laboratory for Biomechanics and Mechanobiology of Ministry of Education, School of Biological Science and Medical Engineering, Beihang University, Beijing 100191, P.R. China
\\
\bf{3} College of Foreign Languages, Jinan University, Guangzhou 510632, P.R. China
\\
$\ast$ E-mail: tfyf@gipe.edu.cn, yubofan@buaa.edu.cn
\end{flushleft}

\section*{Abstract}
There has been a controversy as to whether or not the non-pathological flat foot and high-arched foot have an effect on human walking activities. The 3D foot scanning system was employed to obtain static footprints from subjects adopting a half-weight-bearing stance. Based upon their footprints, the subjects were divided into two groups: the flat-footed and the high-arched. The plantar pressure measurement system was used to measure and record the subjects' successive natural gaits. Two indices were proposed: distribution of vertical ground reaction force (VGRF) of plantar and the rate of the footprint areas. Using these two indices to compare the natural gaits of the two subject groups, we found that (1) in stance phase, there is a significant difference ($p<0.01$) in the distributions of VGRF of plantar; (2) in a stride cycle, there is also a significant difference ($p<0.01$) in the rates of the footprint areas. Our analysis suggests that when walking, the VGRF of the plantar brings greater muscle tension to the flat-footed while a smaller rate of the footprint areas brings greater stability to the high-arched.

\section*{Introduction}
Foot arches are the result of the successive evolution of basic human activities such as walking~\cite{Susman,Wood,Wang}. Among the vertebrates, only humans have foot arches~\cite{Saltzman}. The anatomic structure of one transverse, one medial longitudinal and one lateral longitudinal arch can perform the functions of buffering, amortizing, stabilizing and generating propulsion in human activities~\cite{Saltzman,Chan}. Research into the shape, structure, and function of the foot arch has never ceased.

Differences exist in the shape and structure of each individual's foot arches, which is related to factors such as age and weight~\cite{Lin,Pfeiffer}. The usual methods to collect the foot arch shape include footprinting, X-ray, plantar pressure measurement, laser scanning measurement and MRI scanning~\cite{Chesnin,Gefen,Kanatli,Younger,Chen}. Using indices such as the footprint ratio and foot arch index, we can divide the foot shape into three categories: normal, high-arched and flat. The morphological flat and high-arched foot are asymptomatic. Some research results demonstrate that the non-pathological flatfoot (flexible flatfoot) does not affect one's physiology or quality of life and therefore does not require therapy~\cite{Walczak,Chen2003,Pfeiffer}. Others indicate that the flat foot exerts effects on velocity, stamina and/or balance~\cite{Dowling,Esterman} while those possessing the high-arched foot are unsuitable sprint athletes~\cite{Nattiv}. What difference exists in the gaits of people with the non-pathological flat foot and high-arched foot? Will this difference exert an effect upon walking? If so, how? As yet no satisfactory answers have been provided to these questions.

Shape is a representation of structure. A 3D foot scanner can be used to record the footprint of a person while standing. According to footprint ratio ~\cite{Igbigbi}, flat-footed or high-arched subjects can be selected. Structure affects function. The distribution of plantar pressure can qualitatively reflect such information as the structure and function of the foot as well as the control of the whole body in gait~\cite{Orlin}. System Gait Analysis has been used to measure the natural gaits of both groups of subject. Differences in gait between these two groups have been detected by analyzing both the distributions of VGRF of foot and the rates of the footprint areas.

\section*{Materials and methods}
This study was approved by the Science and Ethics Committee of our institute. Before the experiments, the subjects were informed of the objectives, requirements and procedures of the experiments. All gave informed written consent to participate in the study.

During the selection process, we examined the prospective subjects with help from the Orthopedics Department of our clinic to screen and exclude subjects with pathological flat foot or high-arched foot symptoms such as Talipes calcaneovalgus, Congenital talipes equinovarus (CTEV) (club foot), or planter flexion anomaly.

The subjects were asked to stand barefooted after both feet had been sterilized with $75\%$ ethyl alcohol, and their footprints of half weight were captured with a 3D laser scanner.

The footprint-ratio method was used to analyze the foot shape. For the inner side of the podogram, a tangent was drawn from the heel to the inner edge of the metatarsophalangeal joint to measure the widest distance $AB$ of the hollow area of the footprint and the distance ab between this line and the edge of the outer side of the foot. The width of the solid area was bc. The value of $ab/bc$ was calculated. When $bc=0$, let $ab/bc=1$. When $ab/bc\geq0.786$, it was considered to be high-arched foot; when $ab/bc\leq0.258$, a flat foot. Twelve subjects for each group (flat foot and high-arched foot) were chosen (6 male and 6 female for each group). Their foot shape results are shown in Table~\ref{t1}.
\begin{table}
\caption{\label{t1}Basic information of subjects' feet}
\begin{center}
\item[]\begin{tabular}{@{}*{4}{l}}
\hline
Item (Unit) &High-arched foot &Flat foot &Significance\\
\hline
Foot length (mm)&$241.025\pm11.664$&$243.217\pm13.486$&$$\\
Heel breadth (mm)&$60.358\pm4.762$&$62.950\pm2.424$&$$\\
Foot breadth (mm)&$95.783\pm5.826$&$91.283\pm6.025$&$$\\
Height of Instep (mm)&$62.075\pm3.678$&$57.450\pm3.963$&$p<0.05$\\
Foot arch index&$0.257\pm0.007$&$0.237\pm0.016$&$p<0.01$\\
Footprint ratio index&$0.908\pm0.098$&$0.246\pm0.092$&$p<0.01$\\
\hline
\end{tabular}
\end{center}
\end{table}

The experiment started from the subject's standing position (barefooted). After walking two or three steps, they stepped onto a platform. If the first step onto the platform was found to be incomplete or if the subject walked off the platform, or if the gait seemed apparently nonsuccessive, the subject was asked to try again. Data that met our requirements were collected. See Table~\ref{t2}.
\begin{table}
\caption{\label{t2} Basic kinematic parameters of gait}
\begin{center}
\item[]\begin{tabular}{@{}*{3}{l}}
\hline
Item (Unit) &High-arched foot &Flat foot\\
\hline
Step length (cm)&$68.740\pm5.400$&$68.158\pm6.547$\\
Step time (sec)&$0.507\pm0.026$&$0.517\pm0.049$\\
Stance phase (\%)&$60.784\pm1.419$&$61.034\pm2.566$\\
Swing phase (\%)&$39.216\pm1.419$&$38.966\pm2.566$\\
Stride length (cm)&$137.000\pm10.736$&$136.783\pm12.989$\\
Stride time (sec)&$1.014\pm0.044$&$1.034\pm0.095$\\
Cadence (stride/min)&$59.313\pm2.496$&$57.680\pm5.885$\\
Velocity (m/sec)&$4.884\pm0.528$&$4.751\pm0.761$\\
\hline
\end{tabular}
\end{center}
\end{table}

\begin{figure}[!ht]
\begin{center}
\begin{tabular}{cccc}
 \includegraphics[width=10.8cm]{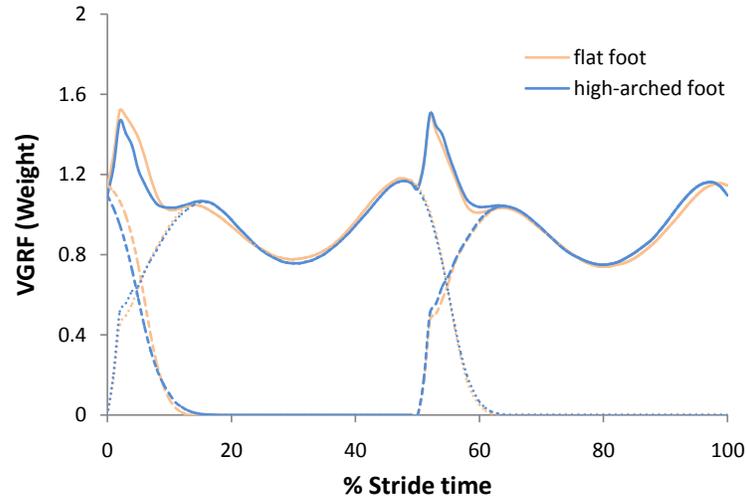}
\end{tabular}
\caption{\label{fig1} Relationships between the forces and time. The blue (deep) dotted lines refer to the VGRFs of the high-arched left and right foot while the blue (deep) solid line refers to the resultant force of the high-arched. The red (light) dotted lines stand for the VGRFs of the flat left/right foot while the red (light) solid line for the resultant force of the flat foot. The VGRFs were obtained from the test report of Zebris FDM, the resultant force from $F_{sum}(t)=F_{left}(t)+F_{right}(t+t_{o})$. Based upon the least-action principle in gait, $t_{o}=\frac{1}{2}T$, where $T$ is the stride cycle time.}
\end{center}
\end{figure}

The VGRF and the stride cycle from the subject were standardized (their weight was normalized as 1). According to the least-action principle in gait (the time of initial foot contact falls right in the middle of the other foot's stride cycle)~\cite{Fan2009}, the relationships between the forces (VGRFs of left and right foot and their resultant force) and the time were established. See Fig.~\ref{fig1}.

\section*{Results and Discussion}
Table~\ref{t1} shows that there exists significant difference ($p<0.05$) in the height of instep, as well as a significant difference ($p<0.01$) in the foot arch index and in the footprint-ratio index for both groups. Table~\ref{t2} reveals there is no significant difference of foot arch shape in gait parameters such as stride frequency, length or velocity. Fig.~\ref{fig1} indicates a similar distribution of VGRF for both groups. (When VGRF is standardized according to weight, there is a substantial similarity of VGRF distribution between the two groups.) Since the VGRF exerted on the foot affects variations of acceleration, velocity, position and mechanical energy of the body's center of mass vertically~\cite{Fan 2010}, while walking, the center of mass of the two groups shares almost the same kinetic and kinematic characteristics.

Generally speaking, we cannot identify flat foot or high-arched foot from gait parameters. It is justifiable to say that based upon gait parameters such as stride length, frequency and GRF, neither the flat foot nor the high-arched foot lead to negative effects on physiology or living quality.

\begin{figure}[!ht]
\begin{center}
\begin{tabular}{cccc}
 \includegraphics[width=10.8cm]{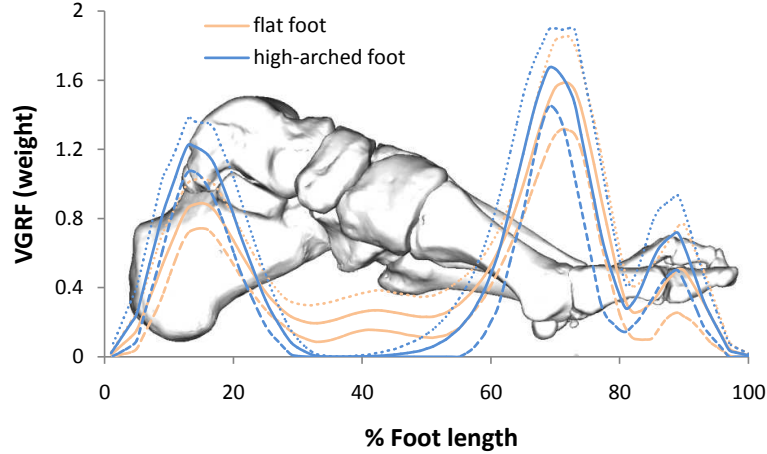}
\end{tabular}
\caption{\label{fig2} Distributions of VGRF of foot in stance phase. The blue (deep) solid line refers to the plantar VGRF of the high-arched foot while the blue (deep) ribbon presents the error bars of the plantar VGRF of the high-arched. The red (light) solid line stands for the plantar VGRF of the flat foot while the red (light) ribbon for the standard error bars of the plantar VGRF of the flat foot. The VGRF has been standardized as 1 by weight, and the stride cycle time and foot length are rated by percentage.}
\end{center}
\end{figure}

We are fully convinced that the principle that structure affects function is truthful. In order to discover the difference between the two foot arch types, we analyzed the distributions of VGRF of foot. The subject's weight, foot length and stance time were standardized. The distributions of VGRF of foot were obtained. See Fig.~\ref{fig2}.

\begin{figure}[!ht]
\begin{center}
\begin{tabular}{cccc}
 \includegraphics[width=10.8cm]{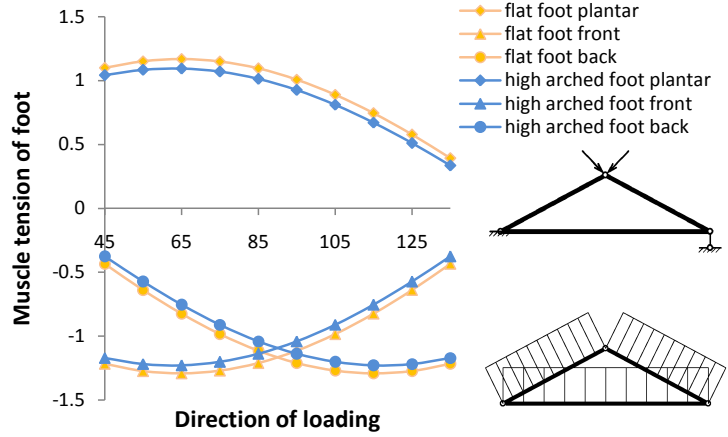}

\end{tabular}
\caption{\label{fig3} Relationship between muscle tension and load. The foot arch is here simplified as a triangle truss, of which the truss width (foot length) is standardized as 1, the truss height (foot arch height) is the foot arch index, and the magnitude of the concentration force is 1. Its direction is changed from $\emph{45}$ degrees to $\emph{135}$ degrees, and its point application is at the top of the truss.}
\end{center}
\end{figure}

Fig.~\ref{fig2} exhibits that for both groups of subjects, significant difference ($p<0.01$) exists in the peak value of VGRF in the heel and center. Virtually no significant difference can be identified under the first metatarsal bone, but significant difference ($p<0.05$) can be noticed under the first proximal phalanx bone. In order to analyze how these differences affect walking, a simplified foot arch structure model (triangle truss) was created by a foot arch index from two types of foot arch. See Fig.~\ref{fig3}.

Fig.~\ref{fig3} indicates that when walking, the structural differences of both foot arch types produce differences in the muscle tension of the foot. When taking a long walk, the flat foot group will feel foot fatigue more easily while the high-arched foot group can walk longer and feel less foot fatigue. While walking, difference exists in the distributions of plantar pressure and the tension that the muscle group under the foot arch bears. This difference is again consistent with the principle that structure affects function.

Stability index plays an important role in gait analysis. Gait stability has been widely discussed by using the symmetry of gait parameters such as the VGRF or the stride length of both types of foot~\cite{Colne}. Fig.~\ref{fig2} shows that the symmetry of gait parameters of both types is very close, indicating the limitation of using such a method to evaluate gait stability.

\begin{figure}[!ht]
\begin{center}
\begin{tabular}{cccc}
 \includegraphics[width=10.8cm]{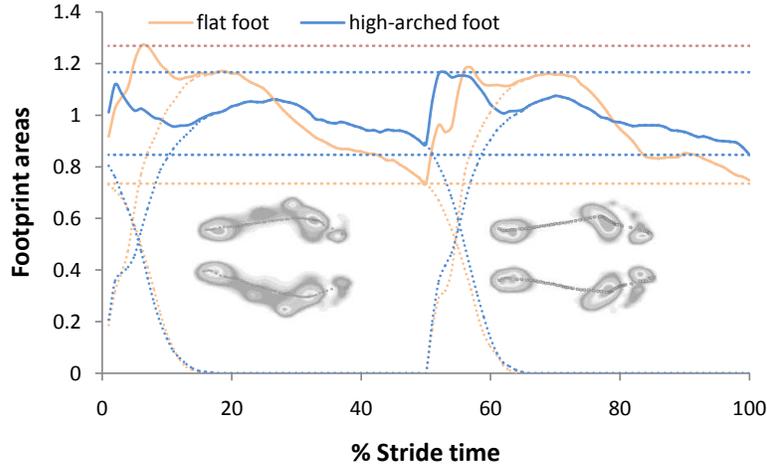}

\end{tabular}
\caption{\label{fig4} Relationship between the footprint areas and stride time. The blue (deep) dotted lines refer to the plantar stance areas of the high-arched left and right foot while the blue (deep) solid line the sum of plantar stance area of the high-arched. The blue (deep) horizontal dotted line presents the variation range of the sum of plantar stance area of the high-arched. The red (light) dotted lines stand for the plantar stance areas of the flat left and right foot while the red (light) solid line for the sum of plantar stance area of the flat-footed. The red (light) horizontal dotted line presents the variation range of the sum of plantar stance area of the flat-footed. The footprint area is derived from the test report of Zebris FDM and the sum of plantar stance area is obtained from $A_{sum}(t)=A_{left}(t)+A_{right}(t+t_{o})$. According to the least-action principle in gait, $t_{o}=\frac{1}{2}T$, $T$ stands for the stride cycle time.}
\end{center}
\end{figure}

Stability analysis of human movement can often be evaluated by the stance area index (for example, the stability is greater when standing on both feet than on one foot.) From this viewpoint, we analyzed the variations of plantar stance area while walking. According to the least-action principle in gait, when the plantar stance area of one stride cycle is standardized, the variation of the plantar stance area of both types of foot can be obtained. See Fig.~\ref{fig4}.
	
Fig.~\ref{fig4} shows that in a stride cycle difference exists in the variations of plantar stance area from both types of subject. In order to evaluate this difference quantitatively, the following equation is applied to assess the rate of plantar stance area while walking:
\begin{equation}\label{eq1}
\sigma=\frac{1}{fT}\sqrt{\sum^{T}_{t=1}\left(A(t)-\bar{A}\right)^{2}}
\end{equation}
where $f$ stands for the collection frequency of the equipment, $T$ the stride cycle time of the subject, $A(t)$ the plantar stance area at a certain moment in a stride cycle and $\bar{A}$ the average value of footprint area in a stride cycle.

Calculation from Eq.~\ref{eq1} provides the variations of footprint areas for the flat-footed and the high-arched groups respectively while walking: $0.147\pm0.041$, $0.084\pm0.034$ ($p<0.01$). The fact that the stability is greater when standing than when walking reveals that the smaller the value from the calculation of Eq.~\ref{eq1}, the better the stability (the minimal value is zero). Accordingly, the rates of footprint areas of both types resulting from Eq.~\ref{eq1} quantitatively describe stability while walking.

\section*{Conclusion}
The structural difference in these types of foot arch causes significant difference of VGRF distribution of foot. The differences in structure and in VGRF distribution have an effect on foot muscle tension while walking. This offers important evidence to analyze foot muscle fatigue. The VGRF distribution of foot can well explain why the flat-footed experience pain more readily when they walk for a long time. The smaller rate of the footprint areas brings greater stability to the high-arched. The lack of stability suffered by the flat-footed requires more consumption of energy, and thus may well explain the fatigue felt by the flat-footed on long walks.

In summary, the mysterious human gait is much more complicated than we had expected. There exist so many unknown phenomena, which we have not yet been able to discover. The establishment of a new gait evaluation index could certainly be employed as an important means to disclose the unknown. The foot arch can not only lessen muscle fatigue, it can also reduce energy consumption~\cite{Collins}. It was the unique foot arch that brought the human being walk out of Africa (so to speak).

How to prevent seniors from falling has always been a key issue in the biomechanical research of gait~\cite{Reid}. The analysis of these two foot types may equally arouse attention to the gait of flat-footed seniors. In addition, does the fact that walking can be affected by foot type mean that we can enhance the gait stability for flat-footed seniors by the design of their shoes~\cite{Mochimaru}? In any case, the VGRF distribution of foot and the rate of footprint areas can be employed as important evaluation indices.

\section*{Acknowledgments}
This project was funded by National Natural Science Foundation of China under the grant $10772053$, $10925208$, $10972061$ and by Key Project of
Natural Science Research of Guangdong Higher Education Grant No $06Z019$. The authors would like to acknowledge the support from the subjects and the clinic of our institute.

\end{document}